\documentclass[11pt,twoside]{article}
\usepackage{CAGN2019}
\usepackage{graphicx}

\usepackage[T1]{fontenc} % Computer Modern (CM) fonts

\usepackage{latexsym}
\usepackage{verbatim}

\usepackage{ifpdf}  
\ifpdf  
      \DeclareGraphicsExtensions{.pdf,.png,.jpg}  
\else  
      \DeclareGraphicsExtensions{.eps}  
\fi 

\newcommand\ion[2]{#1$\;${\scshape{#2}}}% 

\begin{document}
% please do not un-comment the next line
% \input{../../proceeding-book/expages.tex}\setpagenumber{1}

\vskip 1.0cm
\markboth{G.~Dom\'inguez-Guzm\'an et al.}{Chemical abundances and temperature structure of \ion{H}{ii} regions}
\pagestyle{myheadings}
%
%%%%  USE THE LINE THAT DESCRIBES THE CHARACTER OF YOUR WORK %%%%%%
%
\vspace*{0.5cm}
\parindent 0pt{Contributed  Paper}
%\parindent 0pt{Poster}

%\vskip 0.3cm

\vspace*{0.5cm}
\title{Chemical abundances and temperature structure of \ion{H}{ii} regions}

\author{G.~Dom\'inguez-Guzm\'an$^1$, M.~Rodr\'iguez$^1$, C. Esteban$^{2,3}$ and J.~Garc\'ia-Rojas$^{2,3}$}
\affil{
	$^{1}$Instituto Nacional de Astrof\'isica, \'Optica y Electr\'onica,  Apdo. Postal 51 y 216, Puebla, Mexico \\
	$^{2}$Instituto de Astrof\'isica de Canarias, E-38200, La Laguna, Tenerife, Spain \\
	$^{3}$Departamento de Astrof\'isica, Universidad de La Laguna, E-38206, La Laguna, Tenerife, Spain}

\begin{abstract}
We use a sample of 37 \ion{H}{ii} regions with high quality spectra to study the behavior of the relative abundances of several elements as a function of metallicity. The sample includes spectra for eight \ion{H}{ii} regions of the Magellanic Clouds, obtained with UVES/VLT; the rest are gathered from the literature. We find that if we use the traditional two-zone scheme of temperature for the observed ions, the S/O, Cl/O and Ar/O abundance ratios increase with metallicity. However, with slight changes in the temperature structure, which include the use of intermediate temperatures, these ratios are constant with metallicity, as expected. Therefore, high quality observations allow us to deepen our understanding of the temperature structure of \ion{H}{ii} regions.


\end{abstract}
%===================       INTRODUCTION      ======================
\section{Introduction}

The Magellanic Clouds (MCs) are nearby, low-metallicity galaxies, that provide an excellent opportunity to explore in detail the behavior of the chemical composition of \ion{H}{ii} regions as a function of metallicity. However, it is difficult to obtain reliable optical spectra of \ion{H}{ii} regions in the MCs since these galaxies are generally observed at high airmasses, making the observations very sensitive to the effects of atmospheric differential refraction \citep{Fil82}. In fact, of all the available spectra of MC \ion{H}{ii} regions \citep{PT74,D75,PEFW78,DST82,STHM86,TBLDS03,NRMCV03,PA03}, only the spectrum of 30 Doradus presented by \cite{PA03} can be considered of high quality since it is a deep spectrum with high wavelength coverage and relatively high spectral resolution that was obtained using an atmospheric dispersion corrector. Here we present eight new spectra of \ion{H}{ii} regions in the MCs of comparable quality, which, along with a sample of high quality spectra compiled from the literature, allow us to explore different issues related to the analysis of the chemical composition of \ion{H}{ii} regions at different metallicities.

%===================       OBSERVARTIONAL DATA     ======================
\section{Observational data}

\begin{figure}  
	\begin{center}
		\hspace{0.25cm}
		\includegraphics[height=7cm]{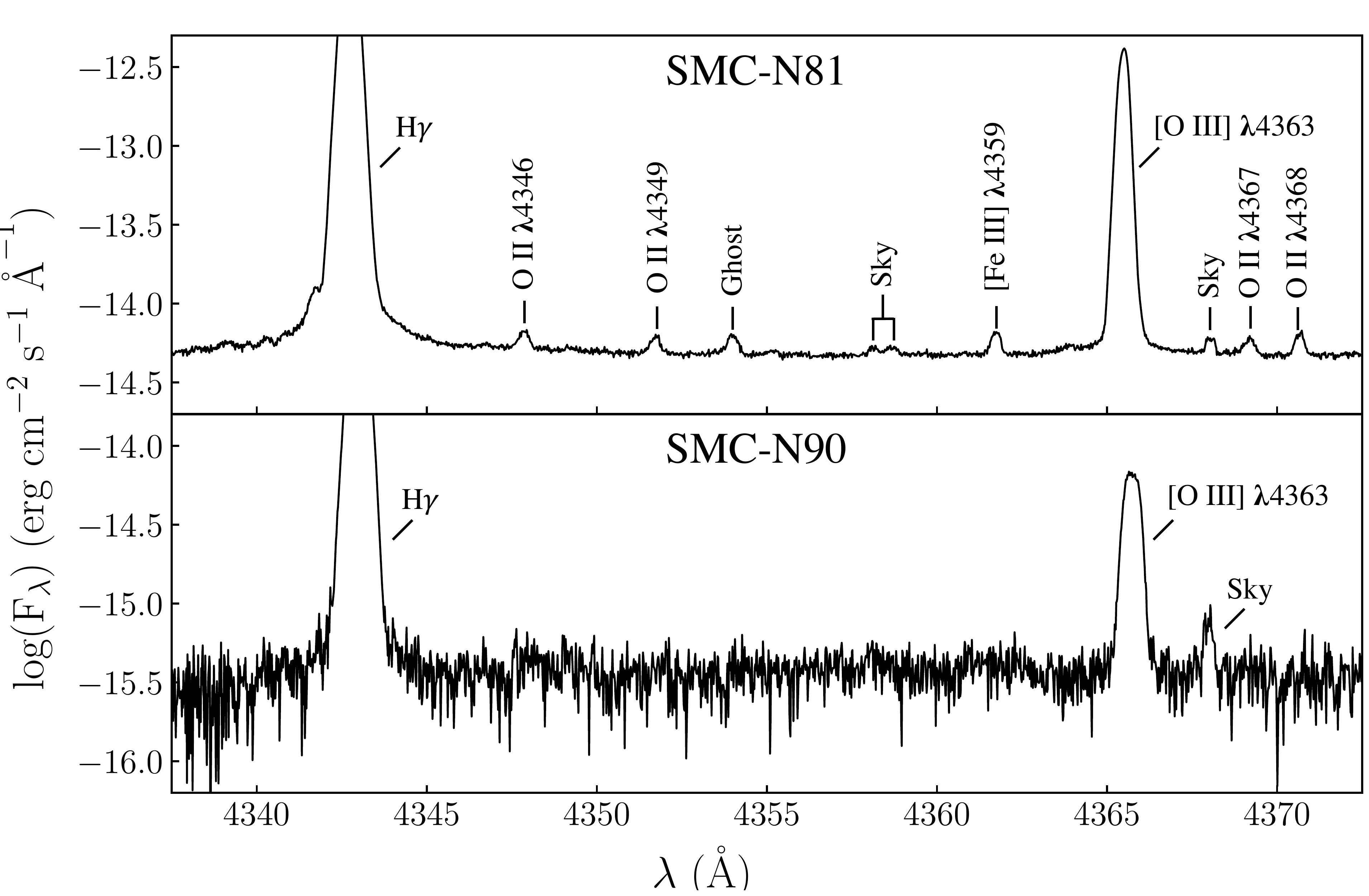}
		\caption{Part of our best (top panel, SMC-N81) and worst (bottom panel, SMC-N90) spectra. The [\ion{O}{iii}] $\lambda 4363$ line used to estimate the electron temperature can be easily measured in both cases.}
		\label{fig1}
	\end{center}
\end{figure}

We obtained echelle spectra of eight \ion{H}{ii} regions in the MCs, four in the Small Magellanic Cloud (SMC)  and four in the Large Magellanic Cloud (LMC). The data were taken with UVES/VLT (Cerro Paranal Observatory, Chile). The spectral range 3100-10400 $\mathring{\mbox{A}}$ was covered with a resolution of $\Delta\lambda \sim \lambda/11600$. The atmospheric dispersion corrector was used to keep the same observed region within the slit at different wavelengths, since the MCs are observed at high airmasses (between 1.35 and 1.88). Figure \ref{fig1} shows the wavelength range that includes the H$\gamma$ and [\ion{O}{iii}]~$\lambda 4363$ lines for our best (SMC-N81) and worst spectrum (SMC-N90).

%-------------  FIGURE 1 -------------

\begin{figure}  
	\begin{center}
		\hspace{0.25cm}
		\includegraphics[height=9.5cm]{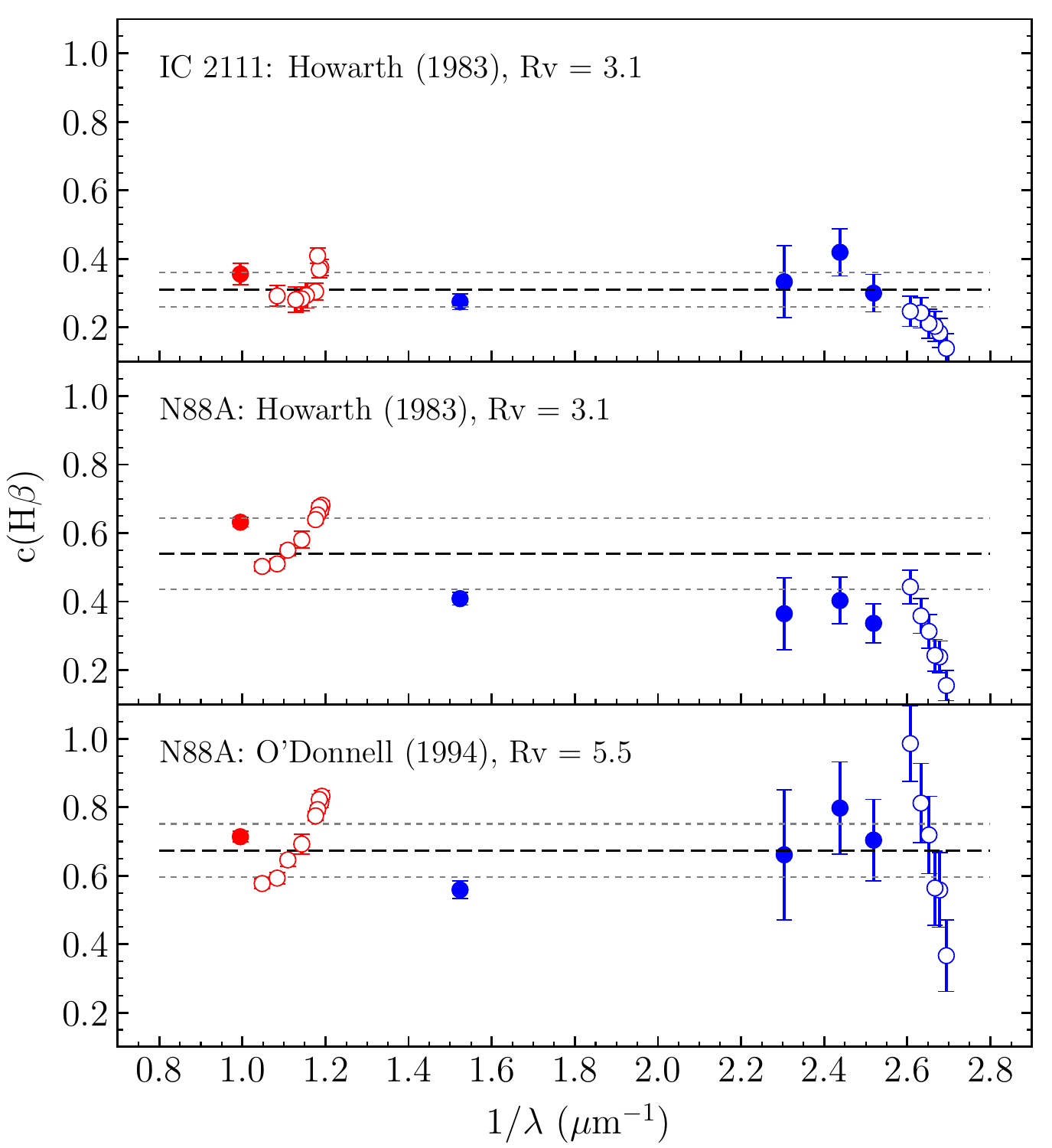}
		\caption{Reddening coefficient $c(\mbox{H}\beta)$ as a function of the inverse wavelength in $\mu$m. Blue circles show the results obtained with the Balmer lines and the red circles those implied by the Paschen lines. The filled circles are those values we use to estimate the weighted mean (long dashed line) and the standard deviation (small dashed lines).}
		\label{fig2}
	\end{center}
\end{figure}

The reddening coefficient, $c(\mbox{H}\beta)$, is determined by comparing the intensities of several Balmer and Paschen lines relative to H$\beta$ with their case~B values \citep{Sto95}. We use lines whose upper levels have principal quantum numbers $n\le7$, since for $n>7$ the lines depart from their expected case~B values (see the empty symbols in Figure~\ref{fig2}). This behavior was previously found in the Orion Nebula by \cite{MD09}, who argue that it could arise from collisions that change the quantum number $l$ by more than $\pm1$ or from the pumping of the \ion{H}{i} lines by absorption of the stellar continuum.

We use the reddening law of \citet{H83}, which has a ratio of total to selective extinction $R_{\mbox{v}}=3.1$ and is commonly used for the MCs, for all objects excepting N88A and N90. In these two regions, the law of \citet{OD94} for $R_{\mbox{v}}=5.5$ provided a better fit. Figure~\ref{fig2} illustrates these results for IC~2111 and N88A.

To this sample of eight MC \ion{H}{ii} regions, we have added 29 \ion{H}{ii} regions from the literature that have spectra of similar quality \citep{GR04,GR05,GR06,GR07,E04,E09,E14,E17,TSC16,PA03}.

%-------------  FIGURE 2 -------------

%===================       RESULTS      ======================

\section{Results}
\label{results}

We have performed a homogeneous analysis of all regions in the sample. The calculations of physical conditions and ionic abundances are carried out with {\sc PyNeb} \citep{LMS15}. We use the density diagnostics $n_{\mbox{\small e}}$([\ion{O}{ii}]), $n_{\mbox{\small e}}$([\ion{S}{ii}]), $n_{\mbox{\small e}}$([\ion{Cl}{iii}]), and $n_{\mbox{\small e}}$([\ion{Ar}{iv}]), and the temperature diagnostics $T_{\mbox{\small e}}$([\ion{N}{ii}]) and $T_{\mbox{\small e}}$([\ion{O}{iii}]) to characterize the gas of the nebula. We estimate the ionic abundances assuming a scheme of two temperatures; $T_{\mbox{\small e}}$([\ion{N}{ii}]) for S$^{+}$, O$^{+}$, and N$^{+}$; and $T_{\mbox{\small e}}$([\ion{O}{iii}]) for S$^{++}$, Cl$^{++}$, O$^{++}$, and Ne$^{++}$. We do not use the other available temperature diagnostics, $T_{\mbox{\small{e}}}$([\ion{O}{ii}]), $T_{\mbox{\small{e}}}$([\ion{S}{iii}]), and $T_{\mbox{\small{e}}}$([\ion{Ar}{iii}]), because they are affected by different problems (recombination effects, telluric absorption) and have higher uncertainties than $T_{\mbox{\small e}}$([\ion{N}{ii}]) and $T_{\mbox{\small e}}$([\ion{O}{iii}]).

The total abundance of nitrogen is obtained using the classical ionization correction factor (ICF), N/O = N$^+$/O$^+$. For the rest of the elements we use the ICFs provided by \citet{DIMS14}. These ICFs were derived using photoionization models tailored for planetary nebulae, but they work better than those provided by \citet{I06} for \ion{H}{ii} regions, leading to somewhat lower dispersions. Besides, the photoionization models for \ion{H}{ii} regions of Reyes-P\'erez et al. (2019, in preparation) confirm their validity.

\begin{figure}  
	\begin{center}
		\hspace{0.25cm}
		\includegraphics[height=8.5cm]{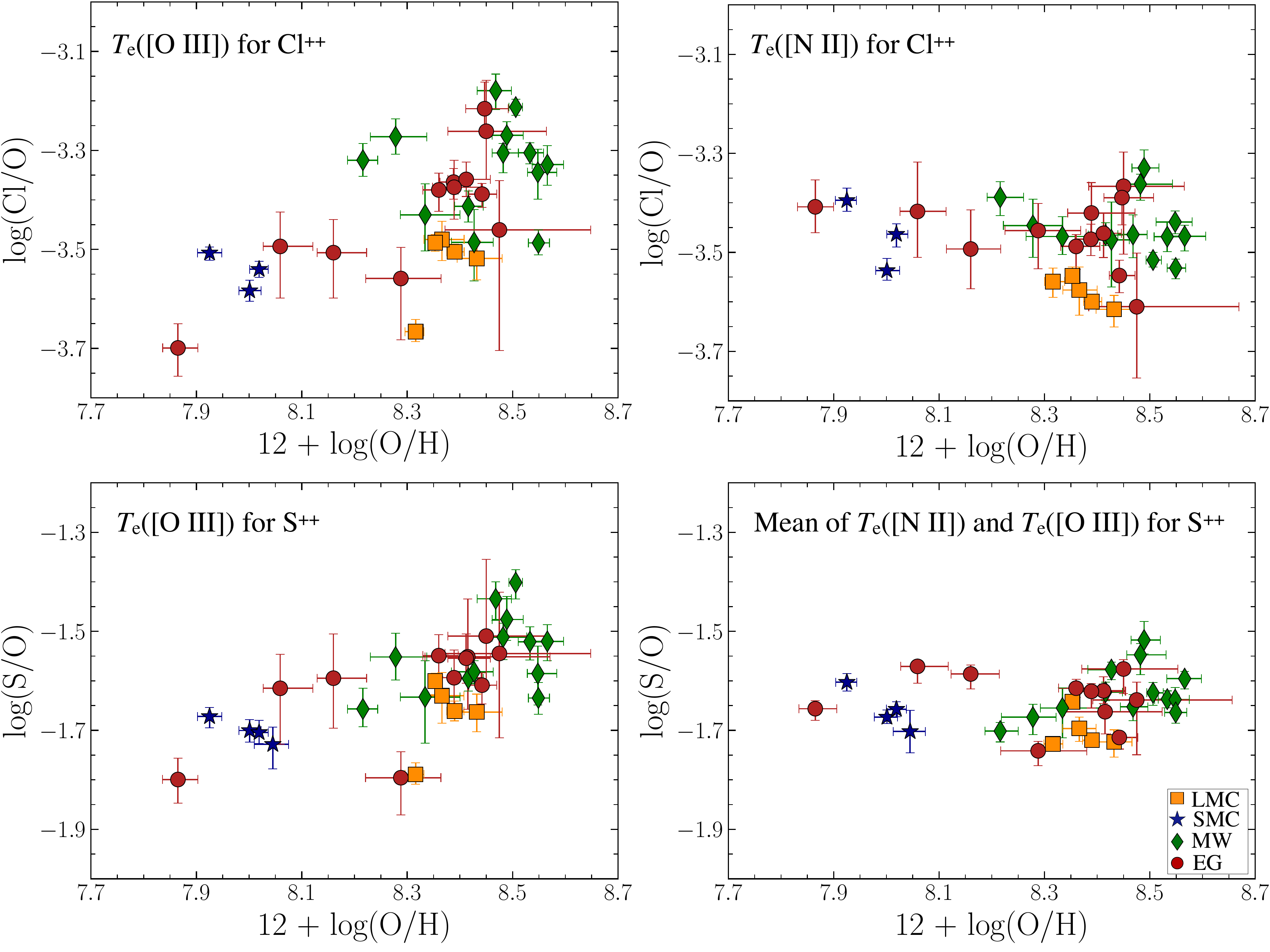}
		\caption{Cl/O and S/O abundance ratios as a function of O/H.  The orange squares are LMC \ion{H}{ii} regions, the blue stars are SMC \ion{H}{ii} regions, the green diamonds are Galactic \ion{H}{ii} regions (MW), and the red circles are extragalactic \ion{H}{ii} regions (EG). The left plots use the traditional method to estimate the total abundance. In the right plots we use $T_{\mbox{\small e}}$([\ion{N}{ii}]) to calculate Cl$^{++}$ and the mean of $T_{\mbox{\small e}}$([\ion{N}{ii}]) and $T_{\mbox{\small e}}$([\ion{O}{iii}]) to determine S$^{++}$.
}
		\label{fig3}
	\end{center}
\end{figure}

\begin{figure}  
	\centering
	\includegraphics[height=4.3cm]{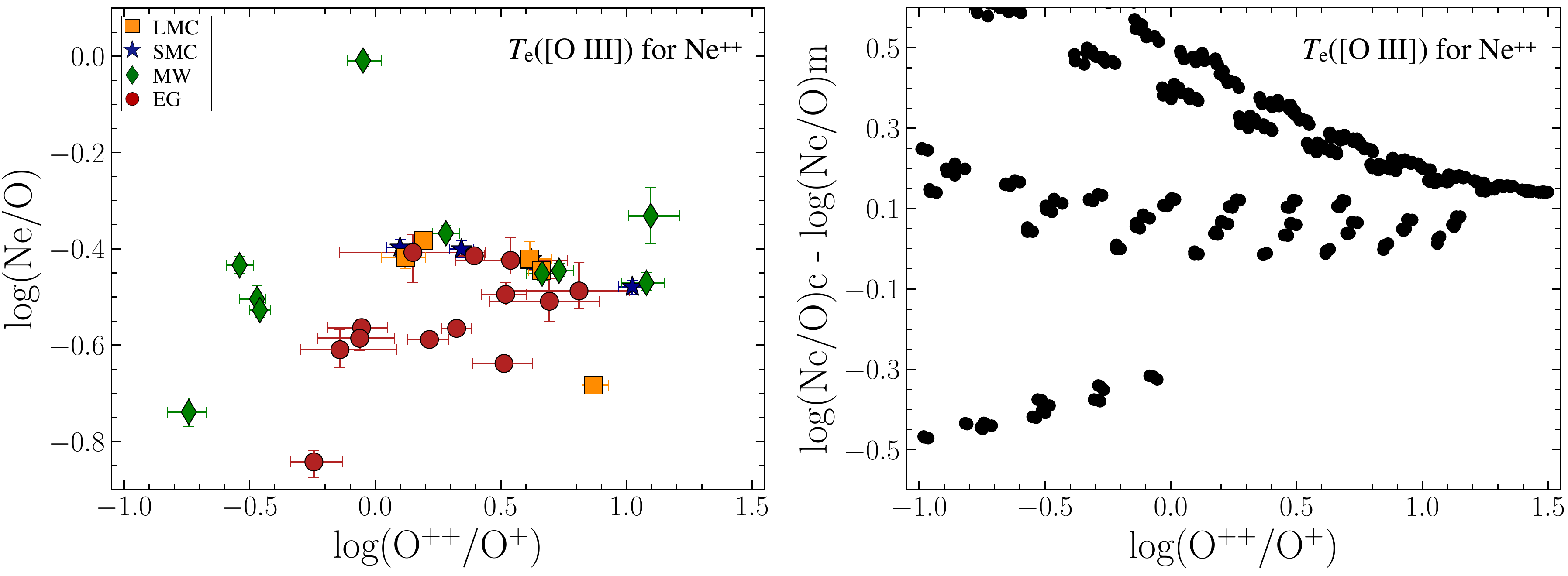}
	\caption{Left panel: Ne/O as a function of the degree of ionization. The color code is the same as in Figure~\ref{fig3}. Right panel: differences between the values of Ne/O obtained from the analysis of the spectra predicted by photoionization models and the input abundance of the models (Reyes-P\'erez et al., 2019, in preparation).
}
	\label{fig4}
\end{figure}

The methodology described above, the most used in the literature, gives us as a result that the Cl/O, Ar/O, and S/O abundance ratios increase with metallicity (left panels in Figure~\ref{fig3} for Cl/O and S/O). On the other hand, if we use $T_{\mbox{\small e}}$([\ion{N}{ii}]) to calculate Cl$^{++}$ and the mean of $T_{\mbox{\small e}}$([\ion{N}{ii}]) and $T_{\mbox{\small e}}$([\ion{O}{iii}]) for S$^{++}$ and Ar$^{++}$, we find that Cl/O, Ar/O, and S/O are approximately constant with metallicity (right panels in Figure~\ref{fig3}), as expected. In the case of Ne$^{++}$, we find that the use of $T_{\mbox{\small e}}$([\ion{O}{iii}]) to calculate its abundance is appropriate, but we find a large dispersion in the values derived for Ne/O (see the left panel in Figure~\ref{fig4}). However, this dispersion is due to the behavior of the ICF for Ne, since it is also found when the spectra predicted by the photoionization models of Reyes-P\'erez et al.\ (2019, in preparation) are analyzed in the same way as the observed spectra (right panel of Figure~\ref{fig4}).

%-------------  FIGURE 3 -------------

%-------------  FIGURE 4 -------------

%===================       CONCLUSIONS      ======================
\section{Conclusions}
\label{discussion}

We present new determinations of chemical abundances of eight \ion{H}{ii} regions in the MCs based on deep echelle spectra taken with UVES/VLT. The quality of the spectra allows us to explore which extinction law is working better for each object. We have extended the sample including spectra with the same quality as the main sample. We find that if we use the traditional two-zone scheme of temperature, the abundance ratios Cl/O, S/O and Ar/O increase with metallicity. However, with slight changes in the temperature structure, these abundance ratios are constant with metallicity, as expected.

\acknowledgments We acknowledge support from Mexican CONACYT grant CB-2014-240562 and from MINECO under grant AYA2015-65205-P. G.D.-G. acknowledges support from CONACYT grant 297932.

\bibliographystyle{aaabib}
\bibliography{DominguezGuzman}

\begin{thebibliography}{}

\bibitem[\protect\astroncite{{Delgado-Inglada} et~al.}{2014}]{DIMS14}
{Delgado-Inglada} G., {Morisset} C., {Stasi{\'n}ska} G., 2014,
\newblock {\em \mnras}, {\bf 440}, 536

\bibitem[\protect\astroncite{{Dufour}}{1975}]{D75}
{Dufour} R.~J., 1975,
\newblock {\em \apj}, {\bf 195}, 315

\bibitem[\protect\astroncite{{Dufour} et~al.}{1982}]{DST82}
{Dufour} R.~J., {Shields} G.~A., {Talbot} Jr. R.~J., 1982,
\newblock {\em \apj}, {\bf 252}, 461

\bibitem[\protect\astroncite{{Esteban} et~al.}{2009}]{E09}
{Esteban} C., {Bresolin} F., {Peimbert} M., {Garc{\'{\i}}a-Rojas} J.,
  {Peimbert} A., {Mesa-Delgado} A., 2009,
\newblock {\em \apj}, {\bf 700}, 654

\bibitem[\protect\astroncite{{Esteban} et~al.}{2017}]{E17}
{Esteban} C., {Fang} X., {Garc{\'{\i}}a-Rojas} J., {Toribio San Cipriano} L.,
  2017,
\newblock {\em \mnras}, {\bf 471}, 987

\bibitem[\protect\astroncite{{Esteban} et~al.}{2014}]{E14}
{Esteban} C., {Garc{\'{\i}}a-Rojas} J., {Carigi} L., {Peimbert} M., {Bresolin}
  F., {L{\'o}pez-S{\'a}nchez} A.~R., {Mesa-Delgado} A., 2014,
\newblock {\em \mnras}, {\bf 443}, 624

\bibitem[\protect\astroncite{{Esteban} et~al.}{2004}]{E04}
{Esteban} C., {Peimbert} M., {Garc{\'{\i}}a-Rojas} J., {Ruiz} M.~T., {Peimbert}
  A., {Rodr{\'{\i}}guez} M., 2004,
\newblock {\em \mnras}, {\bf 355}, 229

\bibitem[\protect\astroncite{{Filippenko}}{1982}]{Fil82}
{Filippenko} A.~V., 1982,
\newblock {\em \pasp}, {\bf 94}, 715

\bibitem[\protect\astroncite{{Garc{\'{\i}}a-Rojas} et~al.}{2005}]{GR05}
{Garc{\'{\i}}a-Rojas} J., {Esteban} C., {Peimbert} A., {Peimbert} M.,
  {Rodr{\'{\i}}guez} M., {Ruiz} M.~T., 2005,
\newblock {\em \mnras}, {\bf 362}, 301

\bibitem[\protect\astroncite{{Garc{\'{\i}}a-Rojas} et~al.}{2007}]{GR07}
{Garc{\'{\i}}a-Rojas} J., {Esteban} C., {Peimbert} A., {Rodr{\'{\i}}guez} M.,
  {Peimbert} M., {Ruiz} M.~T., 2007,
\newblock {\em \rmxaa}, {\bf 43}, 3

\bibitem[\protect\astroncite{{Garc{\'{\i}}a-Rojas} et~al.}{2006}]{GR06}
{Garc{\'{\i}}a-Rojas} J., {Esteban} C., {Peimbert} M., {Costado} M.~T.,
  {Rodr{\'{\i}}guez} M., {Peimbert} A., {Ruiz} M.~T., 2006,
\newblock {\em \mnras}, {\bf 368}, 253

\bibitem[\protect\astroncite{{Garc{\'{\i}}a-Rojas} et~al.}{2004}]{GR04}
{Garc{\'{\i}}a-Rojas} J., {Esteban} C., {Peimbert} M., {Rodr{\'{\i}}guez} M.,
  {Ruiz} M.~T., {Peimbert} A., 2004,
\newblock {\em \apjs}, {\bf 153}, 501

\bibitem[\protect\astroncite{{Howarth}}{1983}]{H83}
{Howarth} I.~D., 1983,
\newblock {\em \mnras}, {\bf 203}, 301

\bibitem[\protect\astroncite{{Izotov} et~al.}{2006}]{I06}
{Izotov} Y.~I., {Stasi{\'n}ska} G., {Meynet} G., {Guseva} N.~G., {Thuan} T.~X.,
  2006,
\newblock {\em \aap}, {\bf 448}, 955

\bibitem[\protect\astroncite{{Luridiana} et~al.}{2015}]{LMS15}
{Luridiana} V., {Morisset} C., {Shaw} R.~A., 2015,
\newblock {\em \aap}, {\bf 573}, A42

\bibitem[\protect\astroncite{{Mesa-Delgado} et~al.}{2009}]{MD09}
{Mesa-Delgado} A., {Esteban} C., {Garc{\'{\i}}a-Rojas} J., {Luridiana} V.,
  {Bautista} M., {Rodr{\'{\i}}guez} M., {L{\'o}pez-Mart{\'{\i}}n} L.,
  {Peimbert} M., 2009,
\newblock {\em \mnras}, {\bf 395}, 855

\bibitem[\protect\astroncite{{Naz{\'e}} et~al.}{2003}]{NRMCV03}
{Naz{\'e}} Y., {Rauw} G., {Manfroid} J., {Chu} Y.-H., {Vreux} J.-M., 2003,
\newblock {\em \aap}, {\bf 408}, 171

\bibitem[\protect\astroncite{{O'Donnell}}{1994}]{OD94}
{O'Donnell} J.~E., 1994,
\newblock {\em \apj}, {\bf 422}, 158

\bibitem[\protect\astroncite{{Pagel} et~al.}{1978}]{PEFW78}
{Pagel} B.~E.~J., {Edmunds} M.~G., {Fosbury} R.~A.~E., {Webster} B.~L., 1978,
\newblock {\em \mnras}, {\bf 184}, 569

\bibitem[\protect\astroncite{{Peimbert}}{2003}]{PA03}
{Peimbert} A., 2003,
\newblock {\em \apj}, {\bf 584}, 735

\bibitem[\protect\astroncite{{Peimbert} \& {Torres-Peimbert}}{1974}]{PT74}
{Peimbert} M., {Torres-Peimbert} S., 1974,
\newblock {\em \apj}, {\bf 193}, 327

\bibitem[\protect\astroncite{{Stasi{\'n}ska} et~al.}{1986}]{STHM86}
{Stasi{\'n}ska} G., {Testor} G., {Heydari-Malayeri} M., 1986,
\newblock {\em \aap}, {\bf 170}, L4

\bibitem[\protect\astroncite{{Storey} \& {Hummer}}{1995}]{Sto95}
{Storey} P.~J., {Hummer} D.~G., 1995,
\newblock {\em \mnras}, {\bf 272}, 41

\bibitem[\protect\astroncite{{Toribio San Cipriano} et~al.}{2016}]{TSC16}
{Toribio San Cipriano} L., {Garc{\'{\i}}a-Rojas} J., {Esteban} C., {Bresolin}
  F., {Peimbert} M., 2016,
\newblock {\em \mnras}, {\bf 458}, 1866

\bibitem[\protect\astroncite{{Tsamis} et~al.}{2003}]{TBLDS03}
{Tsamis} Y.~G., {Barlow} M.~J., {Liu} X.-W., {Danziger} I.~J., {Storey} P.~J.,
  2003,
\newblock {\em \mnras}, {\bf 338}, 687

\end{thebibliography}

\end{document}